\documentclass[aps,english,showpacs,twocolumn]{revtex4-1}
\usepackage{amssymb}
\usepackage{amsmath}
\usepackage{graphicx}
\usepackage{epsfig}

\begin{document}

\title{A physical interpretation for the non-Hermitian Hamiltonian}
\author{L. Jin, and Z. Song}
\affiliation{School of Physics, Nankai University, Tianjin 300071, China}

\begin{abstract}
We explore a way of finding the link between a non-Hermitian Hamiltonian and
a Hermitian one. Based on the analysis of Bethe Ansatz solutions for a class
of non-Hermitian Hamiltonians and the scattering problems for the
corresponding Hermitian Hamiltonians. It is shown that a scattering state of
an arbitrary Hermitian lattice embedded in a chain as the scattering center
shares the same wave function with the corresponding non-Hermitian tight
binding lattice, which consists of the Hermitian lattice with two additional
on-site complex potentials, no matter the non-Hermitian is broken $\mathcal{%
PT}$ symmetry or even non-$\mathcal{PT}$. An exactly solvable model is
presented to demonstrate the main points of this article.
\end{abstract}

\maketitle



\section{Introduction}

In general, a non-Hermitian Hamiltonian is said to be physical when it can
have an entirely real energy spectrum. Much effort has been devoted to
establish a parity-time ($\mathcal{PT}$) symmetric quantum theory as a
complex extension of the conventional quantum mechanics \cite{Bender
98,Bender 99,Dorey 01,Bender 02,A.M43,A.M,A.M36,Jones} since the seminal
discovery by Bender \cite{Bender 98}. It is found that non-Hermitian
Hamiltonian with simultaneous $\mathcal{PT}$ symmetry has an entirely real
quantum mechanical energy spectrum and has profound theoretical and
methodological implications. Reseaches and findings relevent to the spectra
of the $\mathcal{PT}$\ symmetric systems are presented, such as exceptional
points \cite{EP}, spectral singularities for complex scattering potentials
\cite{AMSS},\ complex crystal and other specific models \cite{LonghiSS} have
been investigated. At the same time the $\mathcal{PT}$\ symmetry is also of
great relevance to the technological applications based on the fact that the
imaginary potential could be realized by complex index in optics \cite%
{Bendix,Joglekar,Keya,YDChong,LonghiLaser}.\ In fact, such $\mathcal{PT}$
optical potentials can be realized through a judicious inclusion of index
guiding and gain/loss regions and the most interesting aspects associated
with $\mathcal{PT}$ symmetric system are observed during dynamic evolution
process \cite{Klaiman,El-Ganainy,Makris,Musslimani}.

Thus one of the ways of extracting the physical meaning of a
pseudo-Hermitian Hamiltonian with a real spectrum is to seek for its
Hermitian counterparts \cite{A.M38,A.M391,A.M392}. The metric-operator
theory outlined in Ref. \cite{A.M} provides a mapping of such a
pseudo-Hermitian Hamiltonian to an equivalent Hermitian Hamiltonian. Thus,
most of the studies focused on the quasi-Hermitian system, or unbroken $%
\mathcal{PT}$ symmetric region. However, the obtained equivalent Hermitian
Hamiltonian is usually quite complicated \cite{A.M,JLPT}, involving
long-range or nonlocal interactions, which is hardly realized in practice.

To anticipate these problems, alternative proposals for the connection
between a pseudo-Hermitian Hamiltonian and a real physics system have been
suggested in the context of scattering problems \cite{JLScat}. Central to
that analysis was the recognition that the $\mathcal{PT}$ Hamiltonian may be
used to depict the resonant scattering for an infinite system. It is shown
that any real-energy eigenstate of certain $\mathcal{PT}$ tight-binding
lattice shares the same wave function with a resonant transmission state of
the corresponding Hermitian lattice. In such a framework, further questions
to ask are whether the requirements of the entireness of the real
eigenvalunes and the $\mathcal{PT}$ symmetry of the non-Hermitian system are
really necessary.

In this paper, we propose a physical interpretation for a general
non-Hermitian Hamiltonian based on the configurations involving an arbitrary
network coupled with the input and output waveguides. Relevant to our
previous discussion is the interpretation of the imagiary potentials. Based
on this, we make a tentative connection between a non-Hermitian system and
the corresponding large Hermitian system. It is shown that for any
scattering state of such a Hermitian system, the wavefunction within the
center lattice always corresponds to the equal energy eigenfunction of the
non-Hermitian Hamiltonian, no matter it is $\mathcal{PT}$ symmetric or not.
Our formalism is generic and is not limited to the pseudo-Hermitian system.

This paper is organized as follows. Section II is the heart of this paper
which presents a formulism to reduce a scattering process of a Hermitionian
system to the eigen problem of the non-Hermitian system. Section III
consists of two exactly solvable examples to illustrate our main idea.
Section IV is the summary and discussion.

\section{Non-Hermitian reduction of a Hermitian system}

A typical scattering tight-binding network is constructed by a
scattering-center network and two semi-infinite chains as the input and
output leads. The well-established Green function technique \cite%
{Datta,YangAs,JLTrans} can be employed to obtain the reflection and
transmission coefficients for a given incoming plane wave. The corresponding
wave function within the scattering center should be obtained via Bethe
ansatz method. In the following we will show that this can be done by
solving a finite non-Hermitian Hamiltonian. In our previous work \cite%
{JLScat}, a $\mathcal{PT}$ symmetric non-Hermitian Hamiltionian has been
connected to a physical system in the manner that any real-energy eigenstate
of a $\mathcal{PT}$ tight-binding lattice with on-site imaginary potentials
shares the same wave function with a resonant transmission state of the
corresponding Hermitian lattice embedded in a chain. The main aim of this
article is to answer the question of whether such a statement still holds
for broken $\mathcal{PT}$ non-Hermitian or non-$\mathcal{PT}$ lattice. In
the following, we will show that a scattering state of the Hermitian system
always has connection to the eigenstate of its\ non-Hermitian reduction. For
a certain incident plane wave, the scattering problem of the whole infinite
Hermitian system can be reduced to the eigen problem of a finite
non-Hermitian system.

\begin{figure}[tbp]
\includegraphics[ bb=36 445 410 750, width=6.0 cm, clip]{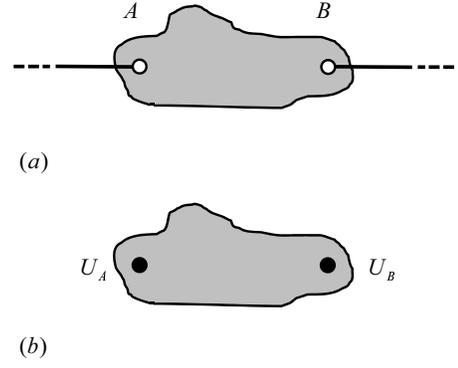}
\caption{Schematic illustration of the configuration of the concerned network.
It consists of an arbitrary graph of a Hermitian tight-binding network
(shadow) connecting to two semi-infinite chains $L$ and $R$ as the waveguides.
The wave function within the scattering center for a scattering state of the
whole system is identical to an equal-energy eigen function of the
non-Hermitian Hamiltonian which is constructed by the center Hermitian
network with imaginary potentials added at the joint sites $A$ and $B$.} %
\label{illus}
\end{figure}

The Hamiltonian of a typical scattering tight-binding network has the form

\begin{equation}
H=H_{A}+H_{B}+H_{c}  \label{H}
\end{equation}%
where
\begin{eqnarray}
H_{A} &=&-J\sum_{i=-1}^{-\infty }b_{i-1}^{\dag }b_{i}-g_{A}b_{-1}^{\dag
}a_{A}+\text{H.c.}  \label{H_a} \\
H_{B} &=&-J\sum_{i=1}^{+\infty }b_{i}^{\dag }b_{i+1}-g_{B}b_{1}^{\dag }a_{B}+%
\text{H.c.}  \label{H_b}
\end{eqnarray}%
represent the left and right waveguides and%
\begin{equation}
H_{c}=-\sum_{i,j=1}^{N}\kappa _{ij}a_{i}^{\dag }a_{j}+\text{H.c.}
\label{H_c}
\end{equation}%
describes an arbitrary $N$-site network as a scattering center. Sites $A$
and $B$ are arbitrary within the network. Here $b_{i}$, $a_{i}$, are boson
(or fermion) operators, $-\kappa _{ii}$ (we denote\ $V_{A}=-\kappa _{AA}$\ $%
V_{B}=-\kappa _{BB}$\ only for the sake of simplicity) represents the
potential at site $i$. Fig. \ref{illus}(a) represents a schematic scattering
configuration for an arbitrary network.


For an incident plane wave incoming from waveguide $A$ with energy $%
E=-2J\cos \left( k\right) $, the scattering wave function can be obtained by
the Bethe ansatz method. The wave function has the form%
\begin{equation}
\left\vert \psi _{k}\right\rangle =\sum_{l}f_{l}b_{l}^{\dag }\left\vert
\text{vac}\right\rangle +\sum_{l}h_{l}a_{l}^{\dag }\left\vert \text{vac}%
\right\rangle  \label{phi_k}
\end{equation}%
where%
\begin{eqnarray}
f_{l} &=&\left\{
\begin{array}{cc}
e^{ik\left( l+1\right) }+re^{-ik\left( l+1\right) }\text{,} & l\in (-\infty
,-1] \\
te^{ik\left( l-1\right) }\text{,} & l\in \lbrack 1,\infty )%
\end{array}%
\right.  \label{f_l} \\
h_{l} &=&h_{l}\text{, }l\in \left[ 1,N\right] .  \notag
\end{eqnarray}%
Here $r$, $t$ are the reflection and transmission coefficients. The explicit
form of the Schr\"{o}dinger equations for the waveguides $H_{A}$ and $H_{B}$
are%
\begin{equation}
H\left\vert \psi _{k}\right\rangle =E\left\vert \psi _{k}\right\rangle
\label{S_eq}
\end{equation}%
admits

\begin{gather}
-Jf_{j-1}-Jf_{j+1}=Ef_{j},  \notag \\
\text{(}j\in (-\infty ,-2]\cup \lbrack 2,+\infty )\text{)}
\label{Leads_Schrodinger_Eqs} \\
-Jf_{-2}-g_{A}h_{A}=Ef_{-1}  \notag \\
-Jf_{2}-g_{B}h_{B}=Ef_{1}.  \notag
\end{gather}%
From Eq. (\ref{Leads_Schrodinger_Eqs}), we obtain $E=-2J\cos \left( k\right)
$\ and

\begin{eqnarray}
h_{A} &=&\frac{J}{g_{A}}\left( e^{ik}+re^{-ik}\right) ,  \label{condition_L}
\\
h_{B} &=&\frac{J}{g_{B}}te^{-ik}.  \label{condition_R}
\end{eqnarray}%
Vanishing $h_{A}$ ($h_{B}$) is beyond of our interest. From Eqs. (\ref%
{condition_L}) and (\ref{condition_R}), one can express the wavefunctions of
two joints ($A$, $B$) as,

\begin{equation}
f_{-1}=\frac{g_{A}}{J}\frac{1+r}{e^{ik}+re^{-ik}}h_{A},  \label{connection_L}
\end{equation}%
\begin{equation}
f_{1}=\frac{g_{B}}{J}e^{ik}h_{B}.  \label{connection_R}
\end{equation}%
The explicit form of the Schr\"{o}dinger equations for $H_{c}$ can be
written as,

\begin{eqnarray}
-\sum_{i}^{N}\kappa _{ij}h_{i} &=&Eh_{j}\text{, (}j\neq A,B\text{)}  \notag
\\
-\sum_{i\neq A}^{N}\kappa _{iA}h_{i}-g_{A}f_{-1} &=&\left( E-V_{A}\right)
h_{A}\text{,}  \label{Seq_H_c} \\
-\sum_{i\neq B}^{N}\kappa _{iB}h_{i}-g_{B}f_{1} &=&\left( E-V_{B}\right)
h_{B}\text{.}  \notag
\end{eqnarray}%
Substituting the expression for $f_{-1}$ and $f_{1}$ from Eqs. (\ref%
{connection_L}) and (\ref{connection_R}), to the above Eqs. (\ref{Seq_H_c}),
we get the following Schr\"{o}dinger equations for the center network,

\begin{eqnarray}
-\sum_{i}^{N}\kappa _{ij}h_{i} &=&Eh_{j}\text{, (}j\neq A,B\text{)}  \notag
\\
-\sum_{i\neq A}^{N}\kappa _{iA}h_{i} &=&\left( E-U_{A}\right) h_{A}\text{,}
\\
-\sum_{i\neq B}^{N}\kappa _{iB}h_{i} &=&\left( E-U_{B}\right) h_{B}\text{.}
\notag
\end{eqnarray}%
with

\begin{eqnarray}
U_{A} &=&V_{A}-\frac{g_{A}^{2}}{J}\frac{1+r}{e^{ik}+re^{-ik}},  \label{U_ab}
\\
U_{B} &=&V_{B}-\frac{g_{B}^{2}}{J}e^{ik}.  \notag
\end{eqnarray}%
This is equivalent to the effective non-Hermitian Hamiltonian%
\begin{equation}
\mathcal{H}=H_{c}+\left( U_{A}-V_{A}\right) n_{A}+\left( U_{B}-V_{B}\right)
n_{B}.  \label{H_eff}
\end{equation}
Without losing generality, we take $r=\left\vert r\right\vert e^{i\delta }$
with $\left\vert r\right\vert <1$. This leads to

\begin{equation}
\text{Im}\left( U_{A}\right) \text{Im}\left( U_{B}\right) =-\frac{\left(
g_{A}g_{B}\sin k\right) ^{2}(1-\left\vert r\right\vert ^{2})}{J^{2}\left[
1+2\left\vert r\right\vert \cos \left( \delta -2k\right) +\left\vert
r\right\vert ^{2}\right] }<0,  \label{Im_U_ab}
\end{equation}%
which means that the imaginary part of the additional potentials have
opposite signs, one providing gain and the other loss. This is in accordance
to the conservation law of the current. It is important to stress that
magnitude of the two imaginary potentials may not equal, which deviates from
the general understanding of an imaginary potential.

The existence of the scattering solution of the Hermitian system $H$ ensures
that there must exist at least one real solution of $\mathcal{H}$\ with
eigenvalue equals to the incident energy $E$.\ It possesses the identical
wavefunction as that of the scattering state within the region of the
scattering center. Then a scattering problem is reduced to the eigen problem
of a non-Hermitian Hamiltonian. This conclusion is an extension of our
previous result \cite{JLScat}. In this work, our formalism is generic: The
central network is not limited to the linear geometry\ and the scattering is
not restricted to be resonant transmission. Thus the scattering
interpretation for the non-Hermitian Hamiltonian is not limited to
pseudo-Hermitian system. This rigorous conclusion has important implications
in both theoretical and methodological aspects.

Likewise, if we consider the inverse scattering process,\textbf{\ }i.e.,
taking the time-reversal operation on the above mentioned scattering
process. The corresponding Bethe ansatz wave function has the form\textbf{\ }%
\begin{equation}
\left\{
\begin{array}{cc}
e^{-ik\left( l+1\right) }+r^{\ast }e^{ik\left( l+1\right) }\text{,} & l\in
(-\infty ,-1] \\
t^{\ast }e^{-ik\left( l-1\right) }\text{,} & l\in \lbrack 1,\infty ) \\
h_{l}^{\ast }\text{, } & l\in \left[ 1,N\right]%
\end{array}%
\right.  \label{inverse}
\end{equation}%
with energy $E=-2J\cos k$. The above conclusion still holds. Straightforward
algebra shows that the corresponding non-Hermitian reduction is $\mathcal{H}%
^{\dag }$. In the framework of non-Hermitian quantum mechanics, $\mathcal{H}%
^{\dag }$\ takes an important role to construct a complete biorthogonal
basis set, which has no physical correspondence. In the context of our
approach, $\mathcal{H}^{\dag }$\ has the same physics as $\mathcal{H}$, in
describing the scattering problem of the same Hermitian system.


\begin{figure}[tbp]
\includegraphics[ bb=0 175 577 618, width=6.5 cm, clip]{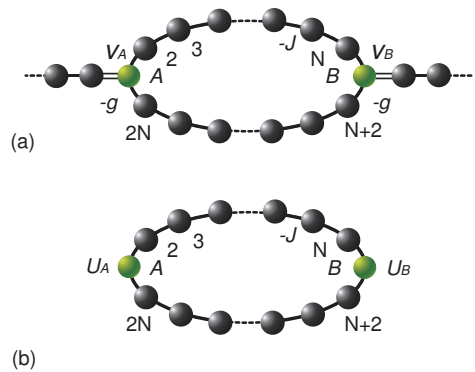}
\caption{(Color online) Schematic illustration of the concrete configuration for a scattering system.
A ring as the scattering center, connects to two semi-infinite chains $L$ and $R$ as waveguides with coupling
$-g$. The on-site potentials at the connections
are $V_A$ and $V_B$. The wave function within the scattering center for a
scattering state of the whole system is identical to an equal-energy
eigen function of the non-Hermitian Hamiltonian which is constructed by the
center Hermitian ring with imaginary potentials $U_A$ and $U_B$ added at the
joint sites $A$ and $B$.} \label{ring}
\end{figure}


\section{ILLUSTRATIVE EXAMPLES}

In this section, we investigate simple exactly solvable systems to
illustrate the main idea of this article. We will discuss two examples which
correspond to a $\mathcal{PT}$ and a non-$\mathcal{PT}$ non-Hermitian
Hamiltonian, respectively. The advangtage of these examples are that the
non-Hermiltian Hamiltonians are exactly solvable.

\subsection{Exactly solvable $\mathcal{PT}$ Hamiltonian}

To exemplify the previously mentioned analysis of relating the stationary
states of a non-Hermitian $\mathcal{PT}$-symmetric Hamiltonian to a
scattering problem for a Hermitian one, we take the center network to be a
simple network: a uniform ring system. We start with the scattering problem
for a class of symmetric systems, the Hamiltonian can be written as

\begin{eqnarray}
&&H_{\text{ss}}=-J\sum_{i=1}^{2N}a_{i}^{\dag }a_{i+1}+\text{H.c.}-\frac{%
\varepsilon }{2}\frac{g^{2}}{J^{2}}\left( n_{1}+n_{N+1}\right)  \notag \\
&&-J\sum_{i=-1}^{-\infty }b_{i-1}^{\dag }b_{i}-J\sum_{i=1}^{+\infty
}b_{i}^{\dag }b_{i+1}+\text{H.c.}  \label{H_ss} \\
&&-gb_{-1}^{\dag }a_{1}-gb_{1}^{\dag }a_{N+1}+\text{H.c.,}  \notag
\end{eqnarray}%
where we denote the connection sites as $a_{A}=a_{1}$ and\ $a_{B}=a_{N+1}$.

The corresponding non-Hermitian Hamiltonian depends on the energy $E$ of the
incident plane wave as well as the parameters $\varepsilon $ and $g$. To be
concise, as an illustrative example, we would like to present the exactly
solvable model, which are helpful to demonstrate our main idea. Therefore,
we will focus on the following configurations:

i) $g\neq \sqrt{2}J$, $E=\varepsilon =\varepsilon _{n}=-2J\cos \left( n\pi
/N\right) $ where $n\in \left[ 1,N-1\right] $. Here we restrict the energy
of the incident plane wave since it will leads to the pure imaginary
potential, thus ensures\ the existence of the exact solution. Straight
forward algebra shows that the problem of solving the Schrodinger equation
is reduced to the eigen problem of the following non-Hermitian Hamiltonian

\begin{equation}
\mathcal{H}^{\left[ n\right] }=-J\sum_{i=1}^{2N}a_{i}^{\dag }a_{i+1}+\text{%
H.c.}+i\gamma _{n}a_{1}^{\dag }a_{1}-i\gamma _{n}a_{N+1}^{\dag }a_{N+1}
\label{Hn}
\end{equation}%
with the imaginary potential%
\begin{equation}
\gamma _{n}=\frac{g^{2}}{J}\sin (\frac{n\pi }{N}).  \label{Gama_n}
\end{equation}%
Obviously, this Hamiltonian depicts a $2N$-site ring with two imaginary
potentials at two symmetrical sites, which is a $\mathcal{PT}$-invariant
Hamiltonian. Note that the magnitude of the imaginary potential is discrete
in order to obtained the exact solutions. In Appendix A, it is shown that
such lattices can be synthesized from the potential-free lattice by the
intertwining operator technique generally employed in supersymmetric quantum
mechanics. The eigen spectrum of $\mathcal{H}^{\left[ n\right] }$\ consists
of%
\begin{eqnarray}
\varepsilon _{j} &=&-2J\cos \left( j\pi /N\right) ,  \label{E_i} \\
\text{( }j &\in &\left[ 1,N-1\right] ,\text{2-fold degeneracy)}  \notag
\end{eqnarray}%
and two additional levels%
\begin{equation}
\varepsilon _{\pm }=\pm \sqrt{4J^{2}-\gamma _{n}^{2}}.  \label{E+/-}
\end{equation}%
The eigenstates with eigenvalue $\varepsilon _{j}$ can be decomposed into
two sets: bonding and antibonding, with respect to the spatial reflection
symmetry about the axis along the waveguides. For the scattering problem,
only the bonding states are involved. It shows that there always exists a
solution in $\left\{ \varepsilon _{j}\right\} $ to match the energy $%
\varepsilon _{n}$ of the incident wave.

From Eqs. (\ref{E_i}, \ref{E+/-}), on can see that a pair of imaginary
eigenvalues appear, i.e., the $\mathcal{PT}$ symmetry is broken when $g>%
\sqrt{2}J$. In general, a non-Hermitian Hamiltonian with a broken $\mathcal{%
PT}$ symmetry is unacceptable because its complex energy eigenvalues make a
hash of the physical interpretation. On the other hand, the $\mathcal{PT}$
symmetry breaking was observed in optics realm experimentally \cite{AGuo}.
In theoretical apects, $\mathcal{PT}$ symmetry in non-Hermitian spin chain
system was discussed \cite{Giorgi}. From the point of view of this article,
we note that even $\mathcal{H}^{\left[ n\right] }$ possesses a broken $%
\mathcal{PT}$ symmetry, the spectrum $\left\{ \varepsilon _{j}\right\} $
still contains the state with the energy $\varepsilon _{j}=\varepsilon _{n}$%
. It is worth mentioning that the broken symmetry does not contradict the
interpretation of the non-Hermitian Hamiltonian (\ref{Hn}). This idencates
that even the $\mathcal{PT}$ symmetry is broken the non-Hermitian
Hamiltonian still has physical significance.

ii) $g=\sqrt{2}J$, $E=\varepsilon \in \left[ -2J,2J\right] $. Here we do not
restrict the energy of the incident plane wave but the magnitue of $g$.
Straight forward algebra shows that the problem of solving the Schrodinger
equation is reduced to the eigen problem of the following non-Hermitian
Hamiltonian

\begin{equation}
\mathcal{H}^{\left[ \varepsilon \right] }=-J\sum_{i=1}^{2N}a_{i}^{\dag
}a_{i+1}+\text{H.c.}+i\gamma _{\varepsilon }a_{1}^{\dag }a_{1}-i\gamma
_{\varepsilon }a_{N+1}^{\dag }a_{N+1}  \label{H_e}
\end{equation}%
with the imaginary potential%
\begin{equation}
\gamma _{\varepsilon }=\sqrt{4J^{2}-\varepsilon ^{2}}.  \label{Gamma_e}
\end{equation}%
From Appendix A, the solution of the Hamiltonian $\mathcal{H}^{\left[
\varepsilon \right] }$ has the same form of Eqs. (\ref{E_i}, \ref{E+/-})\
with $\gamma _{n}$ replaced\ by $\gamma _{\varepsilon }$.\ Here we would
like to see the relation between the Hamiltonians $\mathcal{H}^{\left[ n%
\right] }$\ and $\mathcal{H}^{\left[ \varepsilon \right] }$: Both of them
come from the same model with different coupling constants (with $g\neq
\sqrt{2}J$ and $g=\sqrt{2}J$) and different incident plane waves (with
discrete and continuous spectra). However they have the same structure but
different values of the imaginary potentials. In Appendix A, we provide the
universal solution contains that of $\mathcal{H}^{\left[ n\right] }$\ and $%
\mathcal{H}^{\left[ \varepsilon \right] }$.

Obviously, Hamiltonian $\mathcal{H}^{\left[ \varepsilon \right] }$ is always
exact $\mathcal{PT}$ symmetric. All the eigenvalues are real. Among them we
can find that $\varepsilon _{\pm }=\pm \sqrt{4J^{2}-\gamma _{\varepsilon
}^{2}}$ $=\pm \varepsilon $,\ one of $\varepsilon _{\pm }$ equals to the
energy of incident plane wave $\varepsilon $ and thus verifies the above
mentioned conclusion. Furthemore, the solution of it has the following
peculiar feature: in the case of $\varepsilon =\varepsilon _{n}$,\ i.e., the
incident wave has wave vector $n\pi /N$ $\left( n\in \left[ 1,N-1\right]
\right) $, the exceptional points appear in $\mathcal{H}^{\left[ \varepsilon %
\right] }$. It is shown in Appendix A that the corresponding eigenfunctions
of $\varepsilon _{+}$\ ($\varepsilon _{-}$) and $\varepsilon _{n}$\ ($%
\varepsilon _{N-n}$) coalesce.

According to non-Hermitian quantum mechanics, in general, $\mathcal{H}^{%
\left[ \varepsilon \right] }$\ has the Hermitian counterpart $H^{\left[
\varepsilon \right] }$ which possesses the same spectrum. When the potential
$\gamma _{\varepsilon }$ approches $\gamma _{\varepsilon _{n}}$, the
similarity transform that connects $\mathcal{H}^{\left[ \varepsilon \right]
} $\ and $H^{\left[ \varepsilon \right] }$ becomes singular. The Hamiltonian
$\mathcal{H}^{\left[ \varepsilon \right] }$ becomes a Jordan-block operator,
which is nondiagonalizable and has\ fewer energy eigenstates $(N-1)$ than
eigenvalues $(N+1)$, (i.e., the lack of completeness of the energy
eigenstates.) Such a Hamiltonian has no Hermitian counterpart \cite%
{BenderPRD}. According to our analysis, one can see that even at the
exceptional points \cite{EP}\ the coalescing eigenstates still has physical
significance.

\subsection{Exactly solvable non-$\mathcal{PT}$ Hamiltonian}

Now we turn to exemplify the previously mentioned analysis of relating the
stationary states of a non-Hermitian non-$\mathcal{PT}$-symmetric
Hamiltonian to a scattering problem for a Hermitian one. We still take the
center network as a simple network: a uniform ring system with uniform
coupling but none on-site real potentials. The corresponding Hamiltonian can
be written as

\begin{align}
H_{\text{as}}=& -J\sum_{i=1}^{2N}a_{i}^{\dag }a_{i+1}+\text{H.c.}
\label{H_as} \\
& -J\sum_{i=-1}^{-\infty }b_{i-1}^{\dag }b_{i}-J\sum_{i=1}^{+\infty
}b_{i}^{\dag }b_{i+1}  \notag \\
& -Jb_{-1}^{\dag }a_{1}-Jb_{1}^{\dag }a_{N+1}+\text{H.c.}  \notag
\end{align}%
We consider the incident plane wave with energy $E=-2J\cos \vartheta $,
where $\vartheta \in (\pi $, $-\pi )$ without any restriction. Straight
forward algebra shows that the problem of solving the Schrodinger equation
is reduced to the eigen problem of the following non-Hermitian Hamiltonian

\begin{equation}
\mathcal{H}^{\left[ \vartheta \right] }=-J\sum_{i=1}^{2N}a_{i}^{\dag
}a_{i+1}+\text{H.c.}+U_{A}a_{1}^{\dag }a_{1}+U_{B}a_{N+1}^{\dag }a_{N+1},
\label{H_k}
\end{equation}%
where the complex potentials are

\begin{eqnarray}
U_{A} &=&-J\frac{e^{i\vartheta N}\cos \left[ \left( N-1\right) \vartheta %
\right] +i\sin \vartheta }{e^{i\vartheta N}\sin \left[ \left( N-1\right)
\vartheta \right] -\sin \vartheta }2\sin \vartheta ,  \label{U_RL} \\
U_{B} &=&-Je^{i\vartheta }.  \notag
\end{eqnarray}%
We can see that, in general,\textbf{\ }Hamiltonian $\mathcal{H}^{\left[
\vartheta \right] }$ is not $\mathcal{PT}$ symmetric, except in some special
cases. It is hardly to get the analytical solution of such a Hamiltonian in
general cases. Fortunately, what we need to do is to prove that the incident
energy $E=-2J\cos \vartheta $ is always one of the eigenvalues of the
Hamiltonian. In fact, in single-particle basis the matrix representation $%
\mathcal{M}^{\left[ \vartheta \right] }$ of the Hamiltonian (\ref{H_k})\
satisfies%
\begin{equation}
\det \left\vert \mathcal{M}^{\left[ \vartheta \right] }+2J\cos \vartheta
\right\vert =0,  \label{det}
\end{equation}%
according to the derivation given in Appendix B. This result do not depend
on the pseudo-Hermiticity of the Hamiltoian. In this sense, one can conclude
that a non-$\mathcal{PT}$ non-Hermitian Hamiltonian still has physical
significance.

\section{Conclusion}

In summary, we have studied the connection between a non-Hermitian system
and the corresponding large Hermitian system. We propose a physical
interpretation for a general non-Hermitian Hamiltonian based on the
configurations involving an arbitrary network coupled with the input and
output waveguides. We employed the Bethe ansatz approach to the scattering
problem to show that for any scattering state of a Hermitian system, the
wavefunction within the scattering center lattice always corresponds to the
equal energy eigenfunction of the non-Hermitian Hamiltonian. It is important
to stress that such a physical interpretation for the non-Hermitian
Hamiltonian is not limited to the pseudo-Hermitian system. As an
application, we examine concrete networks consisting of a ring lattice as
the scattering center. Exact solutions for such types of configurations are
obtained to demonstrate the results. Such results are expected to be
necessary and insightful for the physical significance of the non-Hermitian
Hamiltonian.

\acknowledgments We acknowledge the support of the CNSF (Grant Nos. 10874091
and 2006CB921205).

\appendix

\section{Construction of $\mathcal{PT}$-Hamiltonian by Interwining operator
technique}

In this Appendix, we will derive the central formula for studying the eigen
problem of the $\mathcal{PT}$\ ring system.

\subsection{Linear Transformation}

First of all, the Hamiltonian can be decomposed into two independent
sub-Hamiltonians
\begin{equation}
\mathcal{H}=\mathcal{H}_{\alpha }\mathcal{+H}_{\beta }  \label{HH}
\end{equation}%
\begin{eqnarray}
\mathcal{H}_{\alpha } &=&-J\sum_{i=2}^{N-1}\alpha _{i}^{\dag }\alpha _{i+1}-%
\sqrt{2}J\left( \alpha _{1}^{\dag }\alpha _{2}+\alpha _{N}^{\dag }\alpha
_{N+1}\right) +\text{H.c.}  \notag \\
&&+U_{A}\alpha _{1}^{\dag }\alpha _{1}+U_{B}\alpha _{N+1}^{\dag }\alpha
_{N+1},  \label{H_alpha}
\end{eqnarray}%
\begin{equation}
\mathcal{H}_{\beta }=-J\sum_{i=2}^{N-1}\beta _{i}^{\dag }\beta _{i+1}+\text{%
H.c.}  \label{H_belta}
\end{equation}%
with $\left[ \mathcal{H}_{\alpha },\mathcal{H}_{\beta }\right] =0$, by using
the following linear tranformation:%
\begin{eqnarray}
\alpha _{1} &=&a_{1},\alpha _{N+1}=a_{N+1},  \notag \\
\alpha _{j} &=&\frac{1}{\sqrt{2}}\left( a_{j}+a_{2N+2-j}\right) \text{, }%
j\in \left[ 2,N\right] ,  \label{alpha_belta} \\
\beta _{j} &=&\frac{1}{\sqrt{2}}\left( a_{j}-a_{2N+2-j}\right) \text{, }j\in %
\left[ 2,N\right] .  \notag
\end{eqnarray}%
We will focus on the solution of the Hamiltonian $\mathcal{H}_{\alpha }$.
Typically, the solution can be obtained via Bethe ansatz method as shown in
Ref. \cite{JLTrans}. In this Appendix, we will use the intertwining operator
technique to get the solutions in order to reveal their characteristic
features.


\begin{figure}[tbp]
\includegraphics[ bb=145 145 452 332, width=6.0 cm, clip]{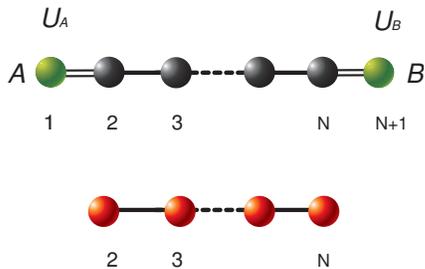}
\caption{(Color online) Schematic illustration of the reduction for a ring
system by linear transformation. The top panel (black) represents the Hamiltonian $\mathcal{H}_{\alpha}$, while the bottom panel (red) shows the Hamiltonian $\mathcal{H}_{\beta}$}%
. \label{scat_half_chain}
\end{figure}


\subsection{Interwining operator technique}

The intertwining operator technique is generally employed in supersymmetric
quantum mechanics, which provides the universal approach to creating new
exactly solvable models. Recently, it is applied to discrete systems in
order to construct the model which supports the desirable spectrum \cite%
{LonghiIOT,LonghiDynamics}.

The critical idea of the intertwining operator technique is as the
following: Consider an $N\times N$ Hamiltonian $H_{1}$ which has the form $%
H_{1}=\mathcal{Q}_{1}\mathcal{R}_{1}+\mu _{1}$, where $\mathcal{Q}_{1}$ and $%
\mathcal{R}_{1}$ represent $N\times (N+1)$ and $(N+1)\times N$ matices,
respectively. One can construct an $(N+1)\times \left( N+1\right) $\ new
Hamiltonian $H_{2}$ ($H_{2}=\mathcal{R}_{1}\mathcal{Q}_{1}+\mu _{1}$) by
interchanging the operators $\mathcal{R}_{1}$ and $\mathcal{Q}_{1}$.\ The
spectrum of $H_{1}$ is the same as that of $H_{2}$ except for the energy
level $\mu _{1}$. Iterating this method results in a series of Hamiltonians $%
H_{3}$, $H_{4}$, $H_{5}$, $\cdots $ whose energy spectra differ from that of
$H_{1}$ owing to the addition of the discrete energy levels $\{\mu _{1},\mu
_{2}\}$, $\{\mu _{1},\mu _{2},\mu _{3}\}$, $\{\mu _{1},\mu _{2},\mu _{3},\mu
_{4}\}$, $\cdots $.

Our aim is to construct a $\mathcal{PT}$-invariant Hamiltonian $H_{3}$ by
adding two energy levels $E=-\mu $ and $E=\mu $ ($0\leqslant \mu \leqslant 2$%
, the obtained conclusion will be extended beyond this region later) into
the energy spectrum of a uniform chain system. We will show the processes of
this construction explicitly. We start with the following $(N-1)\times
\left( N-1\right) $ Hamiltonian

\begin{equation}
H_{1}=-\sum_{n=1}^{N-2}(\left\vert n\right\rangle \left\langle
n+1\right\vert +\text{H.c.})  \label{H_1}
\end{equation}%
which depicts an $\left( N-1\right) $-site uniform chain. The spectrum of $%
H_{1}$\ can be expressed as%
\begin{equation}
\varepsilon _{n}=-2\cos (\frac{n\pi }{N}),n\in \left[ 1,N-1\right] .
\label{eph}
\end{equation}%
On the other hand, $H_{1}$ can be written in the form%
\begin{equation}
H_{1}=\mathcal{QR}-\mu
\end{equation}%
where%
\begin{eqnarray}
\mathcal{Q} &=&\sum_{n=1}^{N-1}(q_{n}\left\vert n\right\rangle \left\langle
n\right\vert +\bar{q}_{n}\left\vert n\right\rangle \left\langle
n+1\right\vert )  \label{QR} \\
\mathcal{R} &=&\sum_{n=1}^{N-1}(r_{n}\left\vert n\right\rangle \left\langle
n\right\vert +\bar{r}_{n}\left\vert n+1\right\rangle \left\langle
n\right\vert )  \notag
\end{eqnarray}%
and%
\begin{equation}
\begin{array}{l}
\mu =2\cos \kappa ,(\kappa >0) \\
r_{n}=q_{n}=-e^{-i\kappa /2} \\
\bar{r}_{n}=\bar{q}_{n}=e^{i\kappa /2}%
\end{array}%
.
\end{equation}%
Then the Hamiltonian $H_{2}$ can be constructed in the form%
\begin{eqnarray}
H_{2} &=&\mathcal{RQ}-\mu  \label{H_2} \\
&=&-\sum_{n=1}^{N-1}(\left\vert n\right\rangle \left\langle n+1\right\vert +%
\text{H.c.})-e^{i\kappa }\left\vert 1\right\rangle \left\langle 1\right\vert
-e^{-i\kappa }\left\vert N\right\rangle \left\langle N\right\vert ,  \notag
\end{eqnarray}%
which possesses an extra eigenvalue $-2\cos \kappa $ based on the spectrum $%
\varepsilon _{n}$.

Next step, we repeat the above procedure based on a new Hamiltonian $%
H_{2}^{\prime }$, which is obtained from $H_{2}$ under parity operation $P$,
i.e.,%
\begin{eqnarray}
H_{2}^{\prime } &=&P^{-1}H_{2}P  \label{H_prim} \\
&=&-\sum_{n=1}^{N-1}(\left\vert n\right\rangle \left\langle n+1\right\vert +%
\text{H.c.})-e^{-i\kappa }\left\vert 1\right\rangle \left\langle
1\right\vert -e^{i\kappa }\left\vert N\right\rangle \left\langle N\right\vert
\notag
\end{eqnarray}%
where%
\begin{equation}
P_{ij}=\delta _{i,N+1-j}  \label{P_ij}
\end{equation}%
is the matix representation of mirror reflection. Note that $H_{2}^{\prime }$
and $H_{2}$\ have identical spectra. Accordingly, $H_{2}^{\prime }$ can be
written as the form%
\begin{equation}
H_{2}^{\prime }=\mathcal{Q}^{\prime }\mathcal{R}^{\prime }+\mu
\end{equation}%
where

\begin{eqnarray}
\mathcal{Q}^{\prime } &=&\sum_{n=1}^{N}(q_{n}^{\prime }\left\vert
n\right\rangle \left\langle n\right\vert +\bar{q}_{n}^{\prime }\left\vert
n\right\rangle \left\langle n+1\right\vert )  \label{QP_prim} \\
\mathcal{R}^{\prime } &=&\sum_{n=1}^{N}(r_{n}^{\prime }\left\vert
n\right\rangle \left\langle n\right\vert +\bar{r}_{n}^{\prime }\left\vert
n+1\right\rangle \left\langle n\right\vert )  \notag
\end{eqnarray}%
and

\begin{equation}
\begin{array}{l}
r_{n}^{\prime }=q_{n}^{\prime }=ie^{-i\kappa /2}, \\
\bar{r}_{n}^{\prime }=\bar{q}_{n}^{\prime }=ie^{i\kappa /2} \\
r_{1}^{\prime }=q_{1}^{\prime }=i\sqrt{2}e^{-i\kappa /2} \\
\bar{r}_{N}^{\prime }=\bar{q}_{N}^{\prime }=i\sqrt{2}e^{i\kappa /2}%
\end{array}%
.
\end{equation}

Finally, the target Hamiltonian $H_{3}$ can be constructed in the form%
\begin{eqnarray}
H_{3} &=&\mathcal{R}^{\prime }\mathcal{Q}^{\prime }+\mu  \label{H_3} \\
&=&-\sum_{2}^{N-1}\left\vert n\right\rangle \left\langle n+1\right\vert -%
\sqrt{2}\left\vert 1\right\rangle \left\langle 2\right\vert -\sqrt{2}%
\left\vert N\right\rangle \left\langle N+1\right\vert +\text{H.c.}  \notag \\
&&+2i\sin \kappa \left( \left\vert 1\right\rangle \left\langle 1\right\vert
-\left\vert N+1\right\rangle \left\langle N+1\right\vert \right) ,  \notag
\end{eqnarray}%
and the energy spectrum (\ref{E_i}) and (\ref{E+/-}) $\mathcal{H}^{\left[ n%
\right] }$\ and $\mathcal{H}^{\left[ \varepsilon \right] }$can be obtained
by adding the unit $J$.

\subsection{Eigenfunctions of $H_{3}$}

Now we turn to derive the eigen functions of $H_{3}$. The eigenfunctions of
the Hamiltonians $H_{1}$, $H_{2}$ and $H_{3}$ are denoted by $\phi _{n}$, $%
\varphi _{n}$ and $\psi _{n}$, respectively. The eigenfunctions of a uniform
chain can be readily written as

\begin{equation}
\phi _{n}\left( j\right) =\sqrt{\frac{2}{N}}\sin \left( \frac{n\pi j}{N}%
\right) ,\text{(}n\in \lbrack 1,N-1]\text{),}  \label{ESH1}
\end{equation}%
According to the interwining operator technique of supersymmetry theory, we
have%
\begin{eqnarray}
\varphi _{n}\left( j\right) &=&\mathcal{R}\phi _{n},n\in \lbrack 1,N-1]
\label{ESH2} \\
\varphi _{N}\left( j\right) &=&e^{-i\kappa j},j\in \lbrack 1,N],  \notag
\end{eqnarray}%
and%
\begin{eqnarray}
\psi _{n}\left( j\right) &=&\mathcal{R}^{\prime }P\varphi _{n},n\in \lbrack
1,N]  \label{ESH3} \\
\psi _{N+1}\left( j\right) &=&\left( -1\right) ^{j}e^{-i\kappa j}\left(
\begin{array}{c}
\frac{1}{\sqrt{2}},j=1,N+1 \\
1,j\in \lbrack 2,N]%
\end{array}%
\right) .  \notag
\end{eqnarray}%
Note that the eigenfunctions are not normalized.

In the above, we restricted $\mu $\ in the region $\left[ 0\text{, }2\right]
$\ for the purpose of obtaining $H_{3}$\ as a non-Hermitian unbroken $%
\mathcal{PT}$\ symmetric Hamiltonian with imaginary potentials at the edges
in the form of Eq. (\ref{H_3}). However, the obtained result can be extended
beyond the region. Actually, one can simply replace $\kappa $\ by $-i\omega $%
\ in all the expressions. Then one can obtain a $\mathcal{PT}$\ Hermitian
Hamiltonian with two added bound states with energy $\pm \mu ,$\ where $\mu
=2\cosh \omega >2$. On the other hand, if $\kappa $\ is replaced by a
complex number $\kappa =\pi /2-i\omega $, one can obtain a non-Hermitian $%
\mathcal{PT}$\ symmetric Hamiltonian in the broken phase. In this case the
two added eigenstates have pure imaginary eigenvalues $\mu =\pm 2i\sinh
\omega $.

\subsection{Coalescence of eigenstates}

Now we investigate the eigenfunctions in the case of $\kappa =k$ ($k=n\pi
/N,n\in \lbrack 1,N-1]$). In this situation, all the eigenfunction can be
written explicitly as%
\begin{equation}
\left\{
\begin{array}{c}
\psi _{n}\left( 1\right) =-i\sqrt{2}\left( -1\right) ^{n}\sin k \\
\psi _{n}\left( j\right) =-2ie^{-\left( N+1-j\right) ik}\sin k \\
\psi _{n}\left( N+1\right) =-i\sqrt{2}\sin k%
\end{array}%
\right. ,  \label{psi_n}
\end{equation}%
\begin{equation}
\left\{
\begin{array}{c}
\psi _{N-n}\left( 1\right) =i\sqrt{2}\left( -1\right) ^{N}\left( -1\right)
^{n}\sin k\  \\
\psi _{N-n}\left( j\right) =-\left( -1\right) ^{\left( N-j\right)
}2ie^{i\left( N+1-j\right) k}\sin k \\
\psi _{N-n}\left( N+1\right) =i\sqrt{2}\sin k%
\end{array}%
\right. ,  \label{psi_N-n}
\end{equation}%
\begin{equation}
\left\{
\begin{array}{c}
\psi _{N}\left( 1\right) =i\sqrt{2}\left( -1\right) ^{n}e^{-ik/2} \\
\psi _{N}\left( j\right) =2ie^{-\left( N+1-j\right) ik}e^{-ik/2} \\
\psi _{N}\left( N+1\right) =i\sqrt{2}e^{-ik/2}%
\end{array}%
\right. ,  \label{psi_N}
\end{equation}%
\begin{equation}
\left\{
\begin{array}{c}
\psi _{N+1}\left( 1\right) =\left( -1\right) ^{j}e^{-ikj}/\sqrt{2} \\
\psi _{N+1}\left( j\right) =\left( -1\right) ^{j}e^{-ikj} \\
\psi _{N+1}\left( N+1\right) =\left( -1\right) ^{j}e^{-ikj}/\sqrt{2}%
\end{array}%
\right. .  \label{psi_N+1}
\end{equation}%
For odd $N$, we have $\psi _{n}\propto \psi _{N}$\ and $\psi _{N-n}\propto
\psi _{N+1}$, which means the coalescence of eigenstates. Also the norms of
the above four eigenstates vanish. For even $N$, we have the same conclusion
except when\textbf{\ }$n=N/2$\textbf{. }In this case, we have $\psi
_{n}\propto \psi _{N}=\psi _{N-n}\propto \psi _{N+1}$ , which means the
coalescence of the three eigenstates. Also the norms of the above three
eigenstates vanish.

\section{Zero determinant}

In this appendix we will prove the Eq. (\ref{det}). Applying the linear
transformation introduced in Appendix A, the $2N$-dimensional matrix $%
\mathcal{M}^{\left[ \vartheta \right] }$\ can be written in a diagonal block
form, i.e.,

\begin{equation}
\mathcal{M}^{\left[ \vartheta \right] }+2J\cos \vartheta =\left[
\begin{array}{cc}
\mathcal{D} & 0 \\
0 & \mathcal{A}%
\end{array}%
\right]  \label{diag DA}
\end{equation}%
where $\mathcal{D}$\ is $\left( N+1\right) $-dimensional, while $\mathcal{A}$%
\ is $N$-dimensional. Then we have

\begin{equation}
\det \left\vert \mathcal{M}^{\left[ \vartheta \right] }+2J\cos \vartheta
\right\vert =\det \left\vert \mathcal{D}\right\vert \det \left\vert \mathcal{%
A}\right\vert .  \label{det AD}
\end{equation}%
Consider the the $\left( N+1\right) $-dimensional matrix $\mathcal{D}$,
which determinant $D=\det \left\vert \mathcal{D}\right\vert $\ has the form

\begin{equation}
D=\left\vert
\begin{array}{cccccccc}
U_{A}-E & -\sqrt{2} &  &  &  &  &  &  \\
-\sqrt{2} & -E & -1 &  &  &  &  &  \\
& -1 & -E & -1 &  &  &  &  \\
&  & -1 & \ddots & \ddots &  &  &  \\
&  &  & \ddots & \ddots & -1 &  &  \\
&  &  &  & -1 & -E & -1 &  \\
&  &  &  &  & -1 & -E & -\sqrt{2} \\
&  &  &  &  &  & -\sqrt{2} & U_{B}-E%
\end{array}%
\right\vert .  \label{D_D}
\end{equation}%
Using cofactor expansion along the first and last rows, we obtain

\begin{eqnarray}
D &=&\left[ E^{2}+U_{A}U_{B}-E\left( U_{A}+U_{B}\right) -4\right] D_{N-1}
\label{D} \\
&&-2\left( U_{A}+U_{B}\right) D_{N-2},  \notag
\end{eqnarray}%
where $D_{j}$ is the $j\times j$\ determinant

\begin{equation}
D_{j}=\left\vert
\begin{array}{cccccc}
-E & -1 &  &  &  &  \\
-1 & -E & -1 &  &  &  \\
& -1 & \ddots & \ddots &  &  \\
&  & \ddots & \ddots & -1 &  \\
&  &  & -1 & -E & -1 \\
&  &  &  & -1 & -E%
\end{array}%
\right\vert .  \label{D_j_M}
\end{equation}%
Such kinds of determinants follow the recursion formula

\begin{equation}
D_{j}=-ED_{j-1}-D_{j-2},  \label{D_j}
\end{equation}%
which leads to

\begin{equation}
D_{j}=\frac{1-e^{2\left( j+1\right) ik}}{1-e^{2ik}}e^{-jik}\text{, }(j<N).
\label{D_N}
\end{equation}%
Substituting the expressions for $U_{A}$, $U_{B}$, from Eq. (\ref{U_RL}), we
get that $D=0$. Thus, Eq. (\ref{det}) is proved.

\end{document}